# TESTING AND IMPLEMENTATION PROGRESS ON THE ADVANCED PHOTON SOURCE (APS) LINEAR ACCELERATOR (LINAC) HIGH-POWER S-BAND SWITCHING SYSTEM

A.E. Grelick, N. Arnold, S. Berg, D. Dohan, G. Goeppner, Y.W. Kang,
A. Nassiri, S. Pasky, G. Pile, T. Smith, S.J. Stein
Argonne National Laboratory, Argonne, Illinois 60439 USA

*Abstract*

An S-band linear accelerator is the source of particles and the front end of the Advanced Photon Source [1] injector. In addition, it supports a low-energy undulator test line (LEUTL) and drives a free-electron laser (FEL). A waveguide-switching and distribution system is now under construction. The system configuration was revised to be consistent with the recent change to electron-only operation. There are now six modulator-klystron subsystems, two of which are being configured to act as hot spares for two S-band transmitters each, so that no single failure will prevent injector operation. The two subsystems are also used to support additional LEUTL capabilities and off-line testing. Design considerations for the waveguide-switching subsystem, topology selection, control and protection provisions, high-power test results, and current status are described.

## 1 INTRODUCTION

The rf power for the APS linear accelerator [2] is provided by five klystrons (L1 through L5), each of which feeds one linac sector. L1 feeds rf power to a thermionic rf gun via the exhaust of one accelerating structure. L2, L4, and L5 are conventional sectors, each using a SLED cavity assembly [3] to feed four accelerating structures. L3 supplies rf power to the photocathode gun located at the beginning of the linac. For normal storage ring injection operation, L1, L2, L4, and L5 are operated, and for the SASE-FEL operation, all five units are operated. A sixth klystron-modulator system was installed in the linac gallery. Design work is in progress on a waveguide distribution and switching system to allow the third and sixth subsystems to serve as hot spares. This is a change from the original version of the system [4], that would have allowed the sixth subsystem to serve as a hot spare for any of the others. The most critical design issue for this system is waveguide switch reliability at 35 MW-peak power.

## 2 TOPOLOGY CHANGES

The change from positron to electron operation in the APS storage ring, together with LEUTL operating requirements, changed the linac configuration by eliminating the L3 accelerating structure. The L3 klystron therefore became an obvious candidate to be used as a hot spare. The current switching system topology is shown in Figure 1. There are now two separate sections. The first, which covers the guns and lower energy sectors is in the process of being installed. In this low-energy section, the L3 klystron serves as a hot spare for the L1 and L2 klystrons and powers either the photocathode rf gun, to support LEUTL operation, or the gun test room. In the second, or high-energy section, the L6 klystron serves as a hot spare for the L4 and L5 klystrons and powers the test stand for switches and other high-power waveguide components. Implementation of the high-power section has been put on hold pending firm decisions on the likely use of higher power klystrons and additional accelerating structures in order to provide increased energy for LEUTL operation.

## 3 HIGH-POWER COMPONENTS

### 3.1 Waveguide Switches

The waveguide switches must be highly reliable at a peak power of 35 MW. Tests had already confirmed that commercially available, sulfur hexafluoride (SF6) pressurized, WR284 waveguide switches were subject to damage due to breakdown at peak powers greater than 30 MW [5]. Scaling to the same field strength in the larger WR340 waveguide yielded a prediction of operation to 43 MW before having significant breakdown problems. Tests of WR340 waveguide switches were set up an additional time at Stanford Linear Accelerator Center (SLAC) Klystron Microwave Laboratory. The unsuccessful results of the original WR340 waveguide switch tests [5] were traced to the fact that, contrary to our expectations, the purchased WR284-to-WR340 transitions were not tapered transitions. This time, electroformed, tapered transitions were used. WR340 switches, which had been reworked by electropolishing, were operated when pressurized with SF6 at 30 PSIG. The results were consistent with the prediction. Three out of four switches operated at a peak power of 43 MW or greater before repetitive arcing occurred. The fourth switch suffered a severe arc during conditioning and showed a decreased return loss as evidence of degradation. To further maximize high-power reliability, an SF6 conditioner-dehydrator system is being used to supply pressure in the interconnecting waveguides and switches.

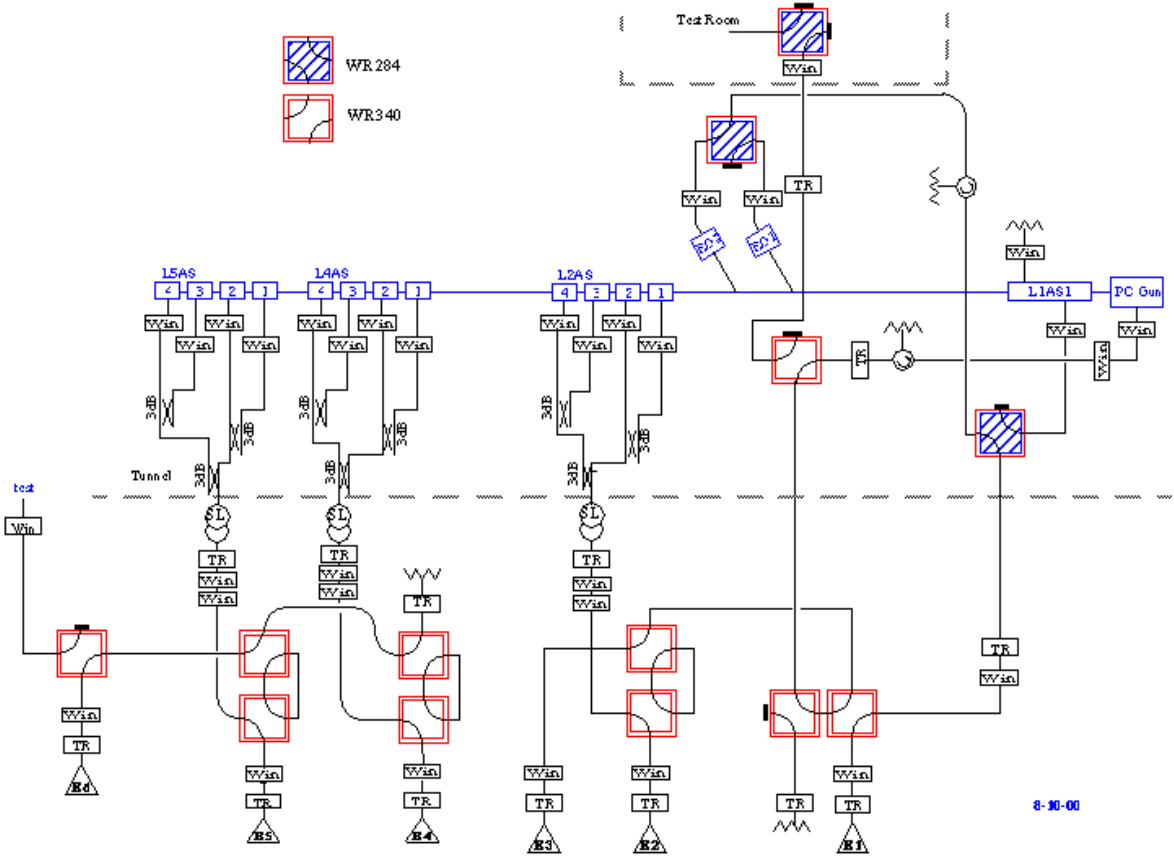

Figure 1: Switching system configuration.

A similar system, which also provides constant circulation of the SF6, has been used successfully used at the Duke University Free-Electron Laser Laboratory at up to 34-MW peak power [6]. Addition of this further refinement has not been implemented to date but remains an option for the future.

### 3.2 New WR340 Window

The WR284 waveguide system uses purchased waveguide windows to provide isolation between vacuum and pressurized sections and to keep to an irreducible minimum any possibility of contamination reaching accelerating structures or SLED cavities.

A window is under development in-house for use with the WR340 waveguide. A return loss of greater than 40 dB on all units has been set as a design goal. The prototype window has achieved a pre-braze measured return loss of 52 dB. However, very strong sensitivity of return loss to assembly pressure has raised concern over the ability to maintain tolerances during the brazing process. Therefore, a modified design, which incorporates tuning adjustments that can be set after brazing, has been created. Figure 2 is a drawing of the modified window design, showing the tuning adjustment provisions.

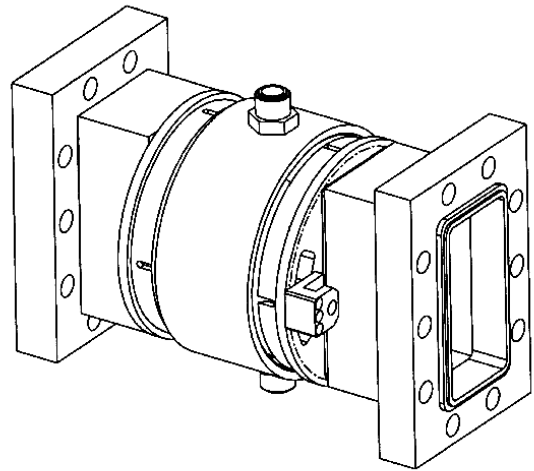

Figure 2: Tunable window design.

## 4 CONTROL IMPLEMENTATION

Implementation of controls for the switching system is being coordinated with the new, more flexible equipment interlock system. Programmable logic controllers (PLCs) were chosen for the new interlocks for their ease of configuration and diagnostic capabilities.

Each sector of the linac will eventually have its own interlock chassis where all interlock cabling is run. The PLC that resides in this chassis (see Figure 3) watches all inputs and generates the proper permits based on the interlock logic. The status of all inputs and outputs from each interlock chassis will be available in the form of operator screens. Additionally, a time-stamped fault stack was created to aid in diagnostics.

The interface to the high-power components of the switching system will utilize the same type of PLC as the new interlock system. This PLC will be responsible only for actuating the switch itself and verifying that the switch is in a valid position. The switching PLC will notify the interlock PLCs via either discreet outputs or high-level messages of the switch positions. The user interface to the switching system will be a rack-mounted touch panel with a key switch. Activating the key switch will disable klystron drive, enable waveguide switch power, and place the switching PLC in the active mode; all accomplished via hard-wired connections.

The design of the interlock logic is complicated by the fact that the cause and effect relationships are dependent on the current operating mode of the linac. For example, if a loss of vacuum in one waveguide section occurs, the proper modulator-klystron must be notified. It is the responsibility of the local PLC interlock system to grant or deny permits based on the input conditions, as well as the switch positions.

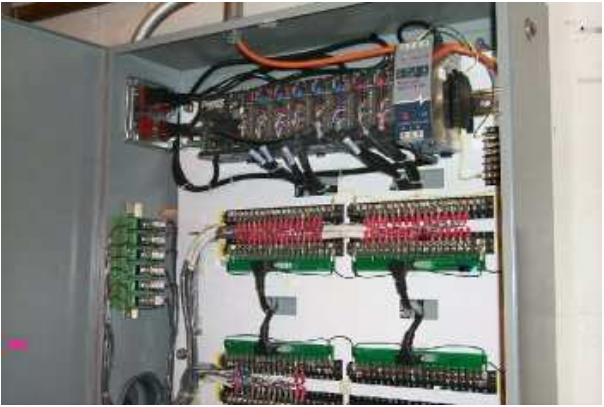

Figure 3: PLC in interlock chassis.

## 5 CURRENT STATUS AND PLANS

Waveguide switch rework and testing is close to completion. The window design is not complete and is the probable gating item in the schedule. Interlock and interface circuits have been tested and are ready to install in most cases. A complete compatibility review is required before the system design can be considered final. In any event, commissioning is expected to be underway during the first half of fiscal year 2001.

## 6 ACKNOWLEDGMENTS

The authors wish to thank G. Caryotakis and S. Gold of SLAC for providing access to their laboratory to perform high-power tests, and J. Eichner and G. Sandoval, also from SLAC, for their help in assembling and operating the test set-up. We also thank D. Meyer for operation of the Argonne high power tests; J. Crandall, H. Deleon, J. Hoyt, T. Jonasson, and J. Warren for setting up the various test configurations; and C. Eyberger for editorial assistance.

This work is supported by the U.S. Department of Energy, Office of Basic Energy Sciences, under Contract No. W-31-109-ENG-38.